\documentclass[prd,aps,amsmath,amssymb,nofootinbib,preprintnumbers]
{revtex4}

\usepackage{graphicx}
\usepackage{dcolumn}
\usepackage{bm}
\usepackage{amsmath}
\usepackage{amsfonts}
\usepackage{ifthen}
\usepackage{psfrag}

\def\ls{\mathrel{\lower4pt\vbox{\lineskip=0pt\baselineskip=0pt
           \hbox{$<$}\hbox{$\sim$}}}}
\def\gs{\mathrel{\lower4pt\vbox{\lineskip=0pt\baselineskip=0pt
           \hbox{$>$}\hbox{$\sim$}}}}
\def\drawbox#1#2{\hrule height#2pt

\hbox{\vrule width#2pt height#1pt \kern#1pt
              \vrule width#2pt}
              \hrule height#2pt}

\def\Asym#1#2{\vcenter{\vbox{\drawbox{#1}{#2}
              \kern-#2pt       
              \drawbox{#1}{#2}}}}

\newcommand{\be}{\begin{equation}}
\newcommand{\ee}{\end{equation}}
\newcommand{\bea}{\begin{eqnarray}}
\newcommand{\eea}{\end{eqnarray}}
\newcommand{\invfb}{\mbox{${\rm fb}^{-1}$}}

\begin{document}
\begin{flushright}
MIFP-09-52
\end{flushright}
\title{Models of supersymmetric dark matter and their predictions in light of CDMS}

\author{Rouzbeh Allahverdi$^{1}$}
\author{Bhaskar Dutta$^{2}$}
\author{Yudi Santoso$^{3}$}

\affiliation{$^{1}$~Department of Physics \& Astronomy, University of New Mexico, Albuquerque, NM 87131, USA \\
$^{2}$~Department of Physics and Astronomy, Texas A\&M University, College Station, TX 77843-4242, USA\\
$^{3}$~Department of Physics and Astronomy, University of Kansas, Lawrence, KS 66045-7582, USA}

\begin{abstract}
We consider the prospects of supersymmetric dark matter in light of the recent results announced by the CDMS experiment. In this paper, we investigate the status of: (i) neutralino dark matter in models of minimal supergravity, (ii) neutralino dark matter in models with nonuniversal Higgs masses, and (iii) sneutrino dark matter in the $U(1)_{B-L}$ extension of the minimal supersymmetric standard model; and discuss the predictions of these models for the LHC, Tevatron, IceCube and PAMELA.
\end{abstract}
\maketitle


\section{Introduction}

There are various lines of evidence for the existence of dark matter in the universe. One proposed solution for the
dark matter problem comes in the form of weakly interacting massive particles (WIMPs) beyond the standard
model~\cite{WIMP}. It is known that for weak scale masses and interactions, thermal freeze out of WIMP annihilation
in the early universe can yield an acceptable relic abundance for dark matter, which is precisely measured by cosmic microwave background (CMB) experiments~\cite{wmap}.

Supersymmetry (SUSY), which is a front-runner candidate to address the hierarchy problem of the standard model (SM), can also provide candidates to explain the dark matter content of the universe. The supergravity motivated (SUGRA) models~\cite{sugra01} have become the focus of major theoretical and experimental activities for the past two decades. The lightest supersymmetric particle (LSP), provided that it is neutral and weakly interacting such as the lightest neutralino, is a natural candidate for thermal dark matter in these models. A simplified version of SUGRA, called minimal supergravity (mSUGRA), which starts
with universal supersymmetry breaking mass parameters at a high scale, has been studied extensively in this context. In
particular, the allowed regions of mSUGRA parameter space that yield acceptable dark matter abundance and are
compatible with all phenomenological constraints have been determined. The prospect for discovery of these regions at
the Large Hadron Collider (LHC) has also been studied~\cite{LHCrelicdensity}. One can go beyond mSUGRA and obtain more realistic scenarios
by relaxing the universality condition in mSUGRA. For example, a partial relaxation is separating the Higgs soft
masses from those of sfermions. The dark matter allowed regions are in general different in these nonuniversal models.

It is also possible to extend the SM gauge symmetries. A minimal extension, motivated by the nonzero neutrino
masses, includes a gauged $U(1)_{B-L}$ symmetry~\cite{mohapatra} ($B$ and $L$ are baryon and
lepton numbers respectively). Anomaly cancellation then implies the existence of three right-handed (RH) neutrinos
and allows us to write the Dirac and Majorana mass terms for the neutrinos to explain the light neutrino masses and
mixings. The $B-L$ extension of the minimal supersymmetric standard model (MSSM) provides new dark matter candidates: the lightest neutralino in the $B-L$ sector~\cite{khalil,ADRS} and the lightest RH sneutrino~\cite{inflation2}. If the $U(1)_{B-L}$ is broken
around TeV, these candidates can acquire the correct thermal relic abundance. They have different qualitative
features from the neutralino dark matter in MSSM: due to the vectorial nature of $B-L$ symmetry there is no
spin-dependent interaction between the dark matter and ordinary matter, and in some regions of the parameter
space the dark matter annihilation rate can get enhancement via the Sommerfeld effect~\cite{ADRS,ADRS2}, which allows an explanation of the recent positron excess in the cosmic ray spectrum as observed by PAMELA~\cite{pamela}. The sneutrino dark matter is particularly interesting because it has a large scattering cross section off nucleons, which is correlated with the mass of $Z^{\prime}$ in the $B-L$ model~\cite{ABDR}. This  $Z'$ can be searched for at the LHC.

There are currently major experimental efforts for detection of dark matter particles. In particular, direct
searches probe the scattering of the dark matter particle off nuclei inside underground detectors. A positive
signal in direct detection experiments would prove the particle nature of the dark matter.
Very recently, results from one underground detection experiment, CDMS, show 2 signal events with $0.6\pm 0.1$ events expected as background~\cite{cdms}. Due to low statistics, this result does not yet provide evidence for dark matter particle, but this can be a hint that the experiment may have started to see something. If these events turn out to be real signals, the direct detection cross section should not be far from the upper limit set by the experiment, $\sim 3.8 \times 10^{-8}$~pb for the dark matter mass around 70~GeV, although there are still some uncertainties from hadronic factors determination, dark halo profile and galactic velocity distribution. Another interpretation of the CDMS result can be seen in ref.~\cite{Bottino:2009km}.
In order to confirm the new findings, this result needs to be validated in the next results of the CDMS experiment (and its upgrades) and other upcoming and future dark matter experiments: XENON 100~\cite{xenon100},  Edelweiss-II~\cite{edelweiss}, CRESST~\cite{cresst}, DEAP~\cite{deap}, CLEAN~\cite{clean}, LUX~\cite{lux} and EURECA~\cite{eureca}.
A large number of similar events in the future will allow us to determine the scattering cross section and also the dark matter
mass more precisely. The results from the LHC and the Tevatron will also be used in tandem in the process of understanding the underlying physics beyond the SM. At this stage it is important to examine the implications of this positive signal for SUSY models that are vigorously searched for in other experiments. We will also examine the impact of using this result as exclusion limit to several SUSY models.

In this paper we discuss the prospects for SUSY dark matter in light of the CDMS results. We consider three scenarios: neutralino dark matter in mSUGRA models, neutralino dark matter in models with nonuniversal Higgs masses, and sneutrino dark matter in the $U(1)_{B-L}$ model, and discuss their predictions for the LHC, Tevatron, IceCube and PAMELA experiments.

\section{mSUGRA Models}

\begin{figure}[ht]
\begin{center}
\includegraphics[width=.48\textwidth]{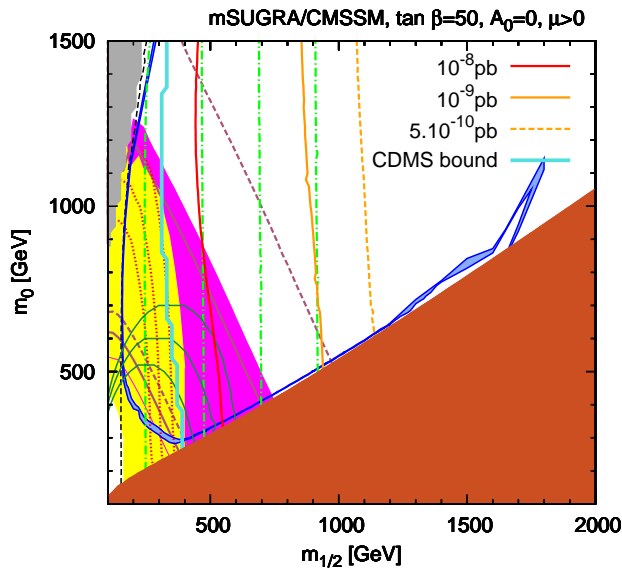}
\end{center}
\vskip -0.25in
\caption{We show the mSUGRA parameter space for $\tan\beta=50$. The contours and shadings are described in the text.}
\label{tanbeta50}
\end{figure}

The mSUGRA model, also known as the constrained minimal supersymmetric standard model (CMSSM), is a simple model that contains only five parameters:
$m_0$ (universal scalar soft mass at $M_{\rm GUT}$), $m_{1/2}$ (universal gaugino mass at $M_{\rm GUT}$), $A_0$ (universal trilinear soft breaking mass at $M_{\rm GUT}$), $\tan \beta = \langle H_1 \rangle /\langle H_2 \rangle$ (where $H_1$ and $H_2$ are the Higgs fields that give rise masses for up- and down-type quarks respectively) and ${\rm sign}(\mu )$ (sign of the Higgs mixing parameter).
We show a typical parameter space in
Fig.~\ref{tanbeta50} for $\tan\beta=$50. The model parameters
are already significantly constrained by different experimental
results. The most important constraints for limiting the parameter space are: (i)~the
light Higgs mass bound of $m_{h^0} > 114.4$~GeV from LEP~\cite{higgs1} (red dotted lines are contours of $m_{h^0} = 112.4$, 113.4 and 114.4~GeV, as calculated using {\tt FeynHiggs-2.6.5}~\cite{FeynHiggs}, respectively from left to right);
(ii)~the $b\rightarrow s \gamma$ branching ratio~\cite{bsgamma}
(95\% CL excluded in the yellow shaded region in Fig.~\ref{tanbeta50}); (iii)~the 2$\sigma$ bound on the dark matter
relic density: $0.106 < \Omega_{\rm CDM} h^2 <0.121$ from WMAP~\cite{wmap} (blue region in Fig.~\ref{tanbeta50});
(iv)~the bound on the lightest chargino mass of $m_{\tilde\chi^\pm_1} >$
103.5~GeV from LEP \cite{aleph} (region left to the black dashed line is excluded) and (v) the muon magnetic moment anomaly
$a_\mu$ (pink shaded region in Fig.~\ref{tanbeta50} is within 2$\sigma$ of where one gets a 3.3$\sigma$ deviation from the SM as suggested by the
experimental results~\cite{amu}). We also show the new 2$\sigma$ contours based on~\cite{Davier:2010rg} by slanted dashed purple lines and  2$\sigma$ contours based on~\cite{Teubner:2010ah}  by slanted solid purple lines. These latest two references use recent changes in the hadronic contribution to calculate the leading order hadronic contribution. Assuming that the future data
confirms the $a_{\mu}$ anomaly, the combined effects of $g_\mu -2$ and
$m_{\tilde\chi^\pm_1} >$ 103.5~GeV then only allows $\mu >0$. The grey shaded region in Fig.~\ref{tanbeta50} is excluded for not satisfying the electroweak symmetry breaking condition, while the red shaded region is excluded because the stau is lighter than the neutralino hence neutralino cannot be the dark matter. The allowed mSUGRA parameter space, selected out by the relic density constraint at present, has five distinct regions~\cite{darkrv}: (i)~the stau-neutralino
($\tilde\tau_1$-$\tilde\chi^0_1$) coannihilation region where
$\tilde{\tau}_1$ is only slightly heavier than $\tilde\chi^0_1$ (which is the LSP), (ii)~the
hyperbolic branch/focus point region where $\tilde\chi^0_1$ has a relatively large Higgsino content, (iii)~the scalar Higgs ($A^0$, $H^0$) annihilation funnel
(2$m_{\tilde\chi^0_1}\simeq m_{A^0,H^0}$),
(iv) the stop-neutralino coannihilation region where the stop particle is almost degenerate with $\tilde\chi^0_1$ when $A_0$ is large,
(v) the bulk
region where none of the properties above is observed, but the relic density is satisfied naturally for small $m_{1/2}$ and $m_0$. The bulk
region is now excluded due to the various existing experimental
bounds. In Fig.~\ref{tanbeta50}, we show the neutralino-proton cross section $\sigma_{\tilde{\chi}_1^0-p}$ contours of $5\times 10^{-10}$, $10^{-9}$ and $10^{-8}$~pb. We also show the CDMS exclusion line, which corresponds to cross section near $5 \times 10^{-8}$~pb, depending on the neutralino mass.  The determination of the neutralino-proton scattering cross section from the experimental results involves hadronic uncertainties (e.g., quarks masses, mass ratios, strangeness content of proton, pion-nucleon sigma term, etc). Therefore, we should allow some room in selecting the parameter space based on the direct detection results. In addition to the constraints above, the branching ratio of the $B_s\rightarrow\mu^-\mu^+$ is also an important constraint for models with large $\tan \beta$.
There has been an active search ongoing at the Tevatron for finding $B_s\rightarrow\mu^- \mu^+$ decay mode. The predictions for the Br$(B_s\rightarrow\mu^- \mu^+)$ are shown in dark green lines: 4.7, 3 and 2$\times 10^{-8}$. The current bound of the branching ratio is Br$(B_s \to \mu^- \mu^+) < 4.7 \times
10^{-8}$~\cite{Aaltonen:2007kv}. We also show the neutralino mass contours (vertical green dot-dashed lines) 100~GeV, 200~GeV, 300~GeV and 400~GeV (respectively from left to right). We see in Fig.~\ref{tanbeta50} that there is a coannihilation region that satisfies all the experimental constraints, including the $g_\mu-2$, with neutralino-proton cross section of order $10^{-8}$~pb and neutralino mass of $\sim 200$~GeV. Fig.~ \ref{tanbeta50} only shows the beginning of the focus point region. In general, the neutralino-proton cross section in the focus point region is larger than $10^{-8}$~pb, although with neutralino masses greater than 100~GeV part of this region is still allowed by CDMS exclusion, especially for larger $m_{1/2}$ and $m_0$.

The solid light-blue line starting at $m_{1/2}=400$ GeV and continues to $m_{1/2}=310$ GeV for $m_0=1500$ GeV shows the CDMS exclusion contour (based on their combined data), excluding the region to the left. The exclusion contour is found to be competitive with the $b\rightarrow s \gamma$ and Br$(B_s\rightarrow\mu^- \mu^+)$ in the stau-neutralino coannihilation region. It is ruling out most of the $g_\mu-2$ favored part of the focus point/hyperbolic region. However, if we allow some room due to the uncertainties, the low focus point region can provide interesting solutions that are compatible with both $g_\mu -2$ and the CDMS two-events. For lower ${\rm tan}\beta$, the CDMS exclusion contour becomes competitive with the Higgs mass constraint from LEP.

\begin{figure}[ht]
\begin{center}
\includegraphics[width=.48\textwidth]{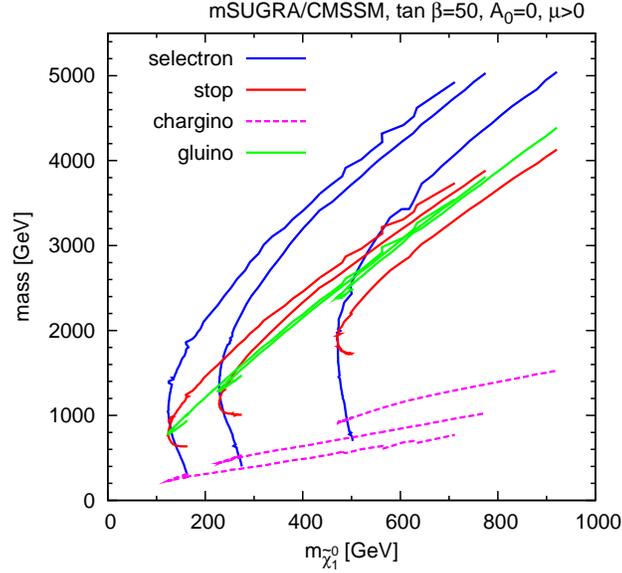}
\end{center}
\vskip -0.28in
\caption{Masses of selectron $\tilde{e}_R$, lightest stop $\tilde{t}_1$, lightest chargino $\tilde{\chi}^\pm_1$ and gluino $\tilde{g}$ as functions of neutralino mass $m_{\tilde\chi^0_1}$ for three fixed values of neutralino proton spin independent elastic scattering cross section of $5 \times 10^{-8}$, $5 \times 10^{-9}$ and $5 \times 10^{-10}$~pb respectively from left to right.}
\label{tb50masses}
\end{figure}

Since the LHC will be searching for the existence of new particles predicted by supersymmetric models, it is interesting to see the range of masses predicted by models with certain values of $\sigma_{\tilde\chi^0_1-p}$, in the  range of $5\times 10^{-10}-5\times 10^{-8}$~pb. In Fig.~\ref{tb50masses} we show the typical mass ranges of the lightest chargino ($\tilde\chi^\pm_1$), the gluino ($\tilde g$), the lightest selectron ($\tilde e_R$) and the lightest stop ($\tilde t_1$) as functions of the lightest neutralino ($\tilde\chi^0_1$) mass for $\tan\beta$=50. We see that at least some sparticles should be  accessible at the LHC if $\sigma_{\tilde\chi^0_1-p} > 10^{-9}$~pb. For example, the chargino, the right selectron, the lighter stop and  the gluino masses are in the range 200~GeV to 1~TeV, 200~GeV to 5~TeV, 500~GeV to 4~TeV and  800~GeV to 4~TeV respectively. The higher values of these masses (e.g., beyond 1~TeV for selectron) appears in the focus point/hyperbolic branch region which is not preferred by the $g_\mu-2$ data. The ranges for these masses do not change much if we change $\tan\beta$. The masses are expected to be measured with high accuracy at the LHC. In fact, it has been analysed for the stau-neutralino coannihilation region~\cite{LHCrelicdensity} that the stau, neutralino and squark masses can be measured at the LHC to a very good accuracy which can be converted to an  uncertainty on $\Omega_{\tilde\chi^0_1} h^2$ of 11~(6)\% using 10~(30) \invfb of data. For the focus point region, the uncertainty becomes 28\% for 300 \invfb. The direct detection cross-section can also be estimated with similar degree of accuracy (e.g., 7\% at 10~(30) \invfb\ for the stau-neutralino coannihilation region) from the LHC measurements, however the main errors in estimating this cross-section arise from the uncertainties of the form factors and the quark masses.

\begin{figure}[ht]
\begin{center}
\includegraphics[width=.48\textwidth]{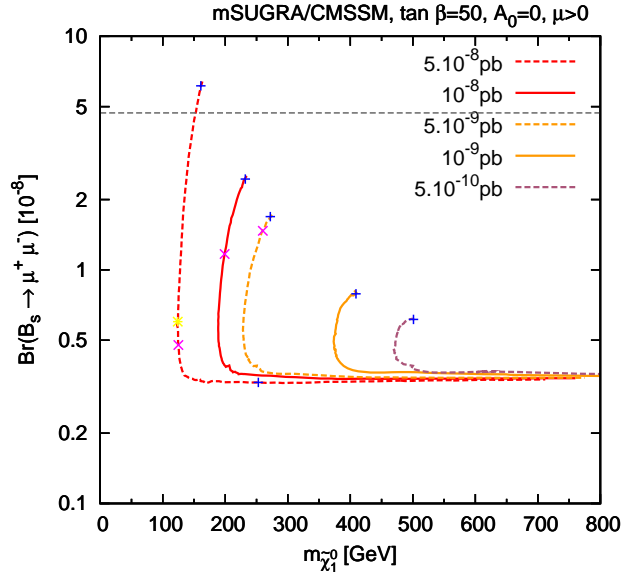}
\end{center}
\vskip -0.25in
\caption{Contours of $\sigma_{\tilde\chi^0_1-p}$ in the plane of Br$(B_s\rightarrow\mu^- \mu^+)$ vs the lightest neutralino mass. The region above the yellow cross is ruled out by $b\rightarrow s\gamma$ constraint, while the regions above  the magenta crosses are favored by the $g_\mu-2$ constraint. The blue crosses show the parameter points favored by WMAP.}
\label{tanbeta50bmm}
\end{figure}

In Fig.~\ref{tanbeta50bmm}, we plot the Br$(B_s\rightarrow\mu^- \mu^+)$ as a function of neutralino mass for five different  $\sigma_{\tilde\chi^0_1 -p}=5\times 10^{-8},\, 1\times 10^{-8},\, 5\times 10^{-9},\, 1\times 10^{-9},\,{\rm and}\, 5\times 10^{-10}$~pb. We find that Br$(B_s\rightarrow\mu^- \mu^+)$ can be probed by the upcoming results at the Fermilab (down to $\sim 2 \times 10^{-8}$ in two years) if the $\sigma_{\tilde\chi^0_1-p}$ is confirmed to be in the order of $10^{-8}$~pb  by the future results of Xenon~100~\cite{xenon100} and CDMS~\cite{cdms} experiments, provided that the muon $g_\mu-2$ anomaly persists. The branching ratio reduces to the SM value $\sim 3\times 10^{-9}$ in the focus point/hyperbolic branch region which appears at large values of $m_0$.

\section{Models with Nonuniversal Higgs Masses}

\begin{figure}[ht]
\begin{center}
\includegraphics[width=.48\textwidth]{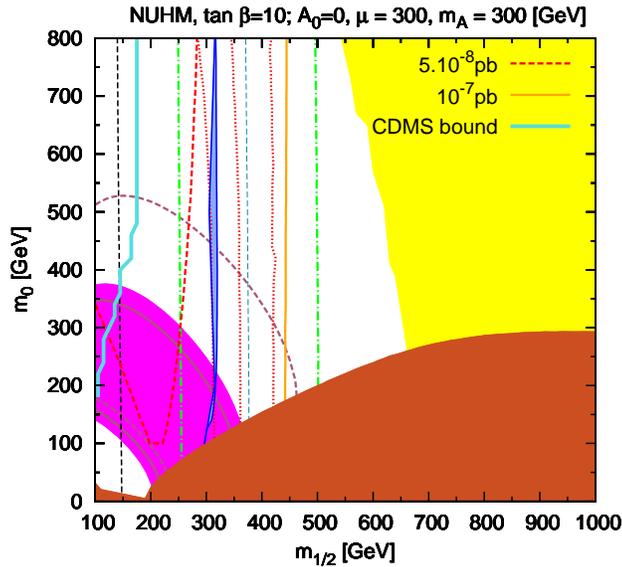}
\end{center}
\vskip -0.25in
\caption{The $m_{1/2}$-$m_0$ plane of NUHM for $\tan \beta = 10$, $A_0 =0$, $m_A=300$~GeV and $\mu=300$~GeV, with contours of $\sigma_{\tilde\chi^0_1-p}$. The  blue region shows the relic density allowed region.}
\label{tanbeta10nuhm}
\end{figure}

We now discuss the models with Nonuniversal Higgs Masses, motivated by the fact that the neutralino-proton cross sections can be enhanced in this model for any value of $m_0$ and $m_{1/2}$ by explicitly modifying the Higgsino content of neutralino.  In the nonuniversal Higgs masses (NUHM) models~\cite{nuhm}, the Higgs soft masses are not tied to $m_0$ at the GUT scale, i.e. $m_{H_1}^2=m_0^2 (1+\delta_1)$ and $m_{H_2}^2=m_0^2 (1+\delta_2)$, where $\delta_1, \delta_2$ are nonuniversal  parameters. The direct detection constraints on the parameter space of these scenarios were discussed in  references~\cite{Higgsnonuni,Higgsnonuni1}. Due to these nonuniversalities at the GUT scale, we can take $\mu$ and $m_A$ to be new parameters at the weak scale. In Fig.~\ref{tanbeta10nuhm}, we show  the contours of $\sigma_{\tilde\chi^0_1-p}$  in the plane of $m_0$ and $m_{1/2}$ for $m_A=300$~GeV and $\mu=300$~GeV. The thin dashed blue line is where $2 m_{\tilde\chi^0_1} = m_A$, and the shaded blue region shows the relic density allowed region due to the neutralino annihilation near the $m_A$ pole.
On the right of the thin blue dashed line, there is further suppression of the neutralino relic density due to the large higgsino content of the neutralino, and also chargino coannihilation effect. Therefore in this region the relic density is always smaller than the WMAP value. If $\mu$ is lowered,  the relic density allowed region will move to the left. As for the mSUGRA case, the red dotted lines from left to right are Higgs masses (112.4, 113.4 and 114.4~GeV), and the black dashed line corresponds to the chargino mass 103.5~GeV. The $\sigma_{\tilde\chi^0_1-p}$ contours are drawn at $5\times10^{-8}$ and $10^{-7}$~pb. The neutralino masses 100~GeV and 200~GeV are shown by green vertical dot-dashed lines. We see that the neutralino proton cross section is getting large when both $\mu$ and $m_A$ are small, and therefore already restricted by the dark matter direct detection searches. The solid light-blue line starting at $m_0=200$~GeV for $m_{1/2}=100$~GeV and continues to $m_{1/2}=170$~GeV for $m_0=800$~GeV shows the  CDMS exclusion contour (based on their combined data). The right side of the contour shows the excluded region. The allowed region increases as we increase $\mu$ and $m_A$. Note however that for the small relic density region, the neutralino proton cross section gets suppression by the ratio of the neutralino relic density over the total dark matter density (WMAP), i.e. assuming a multicomponent dark matter scenario, hence not necessarily excluded.

\begin{figure}[ht]
\begin{center}
\includegraphics[width=.48\textwidth]{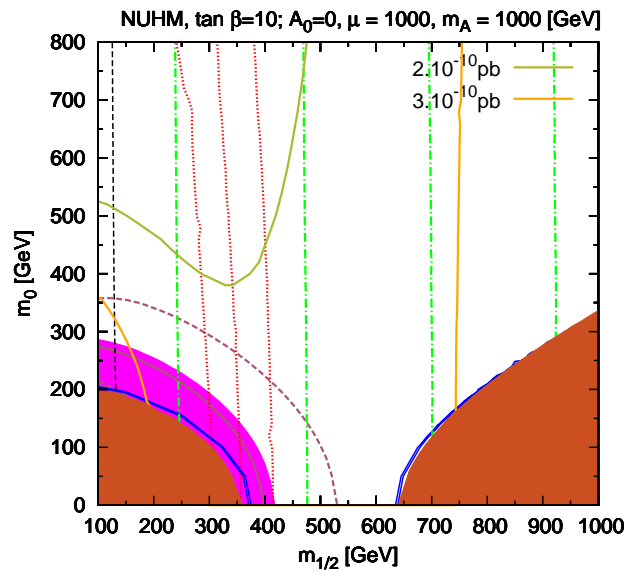}
\end{center}
\vskip -0.25in
\caption{The $m_{1/2}$-$m_0$ plane of NUHM for $\tan \beta = 10$, $A_0 =0$, $m_A=1$~TeV and $\mu=1$~TeV, with contours of $\sigma_{\tilde\chi^0_1-p}$ . The  blue region shows the relic density allowed region.}
\label{tanbeta10nuhm1}
\end{figure}

If we raise $m_A$ and $\mu$ to 1~TeV, as shown Fig.~\ref{tanbeta10nuhm1}, then the higgsino component goes down and consequently, the direct detection cross-section goes down. We find that it is $\sim10^{-10}$~pb for the parameters chosen. Note however that if we raise $\tan\beta$, the cross-section can go up to $10^{-9}$~pb. We also notice that the stau-neutralino coannihilation is the only relic density allowed region shown in this plot. Overall, in the nonuniversal Higgs model, it is possible to find solutions to relic density for any $m_0$ value by tuning the values of $\mu$ and $m_A$. However, intermediate values of $\mu$ are better to explain the cross section emerging from the CDMS data. It is interesting to note that the cross section increases as we increase $m_{1/2}$, in contrast to the mSUGRA case, because of the larger higgsino content in the neutralino. We hope that these nonuniversal models with the intermediate values of $\mu$ can be identified at the LHC. The range of masses for the SUSY spectrum from such a nonuniversal model will be mostly similar to mSUGRA scenarios we presented above for the chosen values of direct detection cross-sections, however, $m_A$  and the neutralino and chargino spectrum will in general be different. Careful measurements of observables (using invariant mass distribution) involving different final states of the SUSY particles will hopefully distinguish these models at the LHC. The entire region is allowed by the CDMS exclusion line which we do not see in this figure, although the cross section might be to low to explain the CDMS two-events. Just like mSUGRA model, the Br($B_s \rightarrow\mu^- \mu^+$) also gets enhanced in the NUHM model for large $\tan \beta$.

\section{$U(1)_{B-L}$ Model}

The minimal $B-L$ model contains a new gauge boson $Z^{\prime}$, two new Higgs fields $H^{\prime}_1$ and $H^{\prime}_2$, the RH neutrinos $N$, and their SUSY partners. The superpotential is
\begin{equation} \label{sup}
W = W_{\rm MSSM} + W_{B-L} + y_D N^c H_1 L \, ,
\end{equation}
where
$L$ denotes the superfield containing the left-handed (LH) leptons (for simplicity, we have omitted the family indices). The $W_{B-L}$ term contains ${H^{\prime}_1},~{H^{\prime}_2}$ and $N^c$ and its detailed form depends on the charge assignments of the new Higgs fields. The last term on the RH side of Eq.~(\ref{sup}) is the neutrino Yukawa coupling term.
Various $B-L$ charge assignments are allowed by anomaly cancellation. We choose the $U(1)_{B-L}$ charges to be $+1$, $1/3$, $+2$ and $-2$ for leptons, quarks, $H^{\prime}_1$ and $H^{\prime}_2$ respectively. Then $H^{\prime}_2$ can couple to the RH neutrinos and give rise to a Majorana mass upon spontaneous breakdown of the $U(1)_{B-L}$.
Choosing these Majorana masses in the $100~{\rm GeV}-1$ TeV range, we have three (dominantly RH) heavy neutrinos and three (dominantly LH) light neutrinos. The masses of the light neutrinos are obtained via the see-saw mechanism.

This model provides
new dark matter candidates: the lightest RH sneutrino ${\tilde N}$~\cite{inflation2} and the lightest neutralino in the $B-L$ sector ${\tilde \chi}^{0 \prime}_1$~\cite{khalil,ADRS}.
In the latter case
the direct detection cross section is almost negligible. We therefore focus here on the RH sneutrino as the dark matter candidate~\footnote{The sneutrino can also lead to successful inflation in the context of the $U(1)_{B-L}$ model~\cite{inflation2}. The dark matter candidate (the RH sneutrino) can then become a part of the inflaton field and thereby gives rise to a unified picture of dark matter, inflation and the origin of neutrino masses. \\
For sneutrino dark matter without $B-L$ see e.g.~\cite{fornengo}.}. It
is made stable by invoking a discrete $R$-parity, but in the context of a $B-L$ symmetry, a discrete matter parity can arise once the $U(1)_{B-L}$ is spontaneously broken~\cite{rparity}. The $B-L$ gauge interactions can yield the correct relic abundance of sneutrinos if the $U(1)_{B-L}$ is broken around the TeV scale.

\begin{figure}[ht]
\begin{center}
\includegraphics[width=8.0cm]{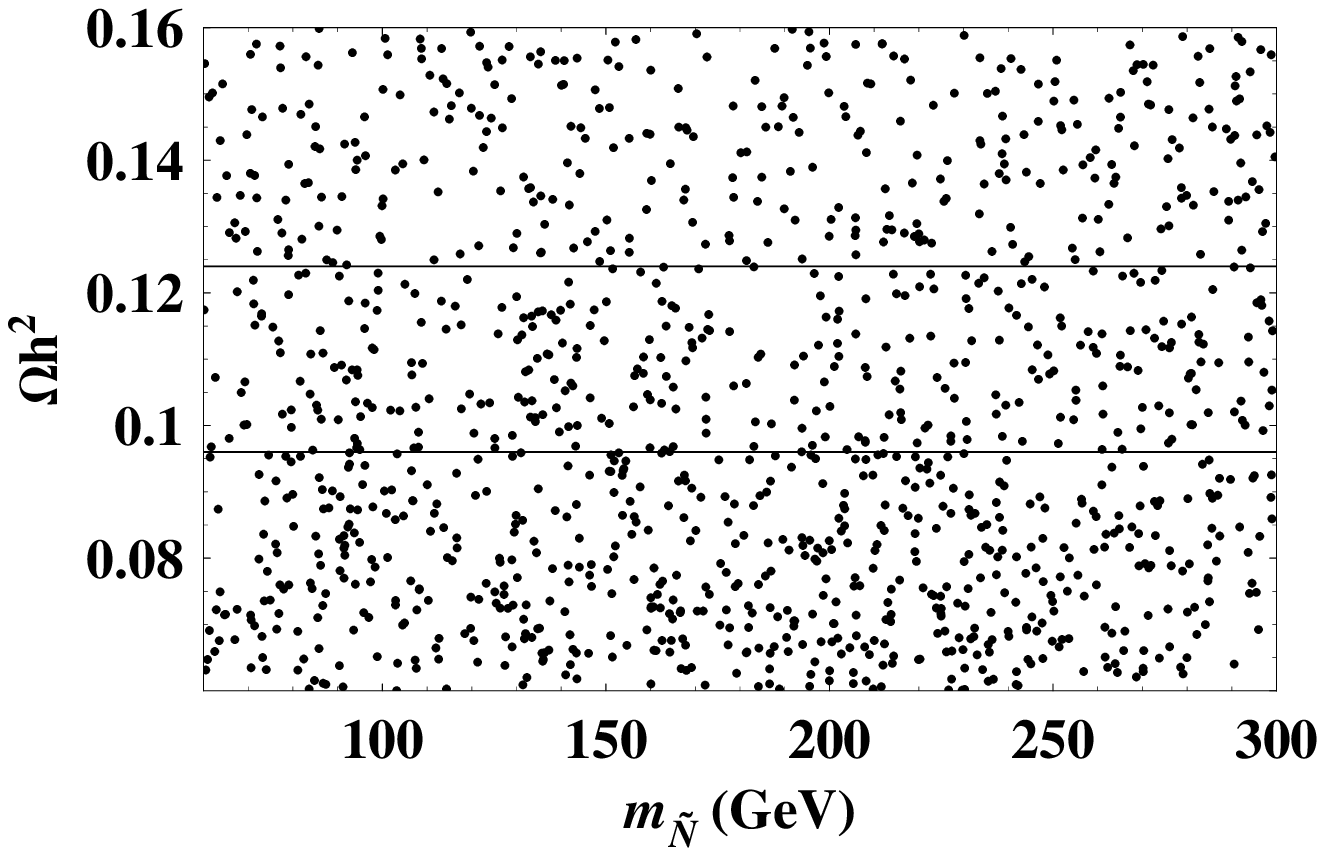}
\includegraphics[width=8.0cm]{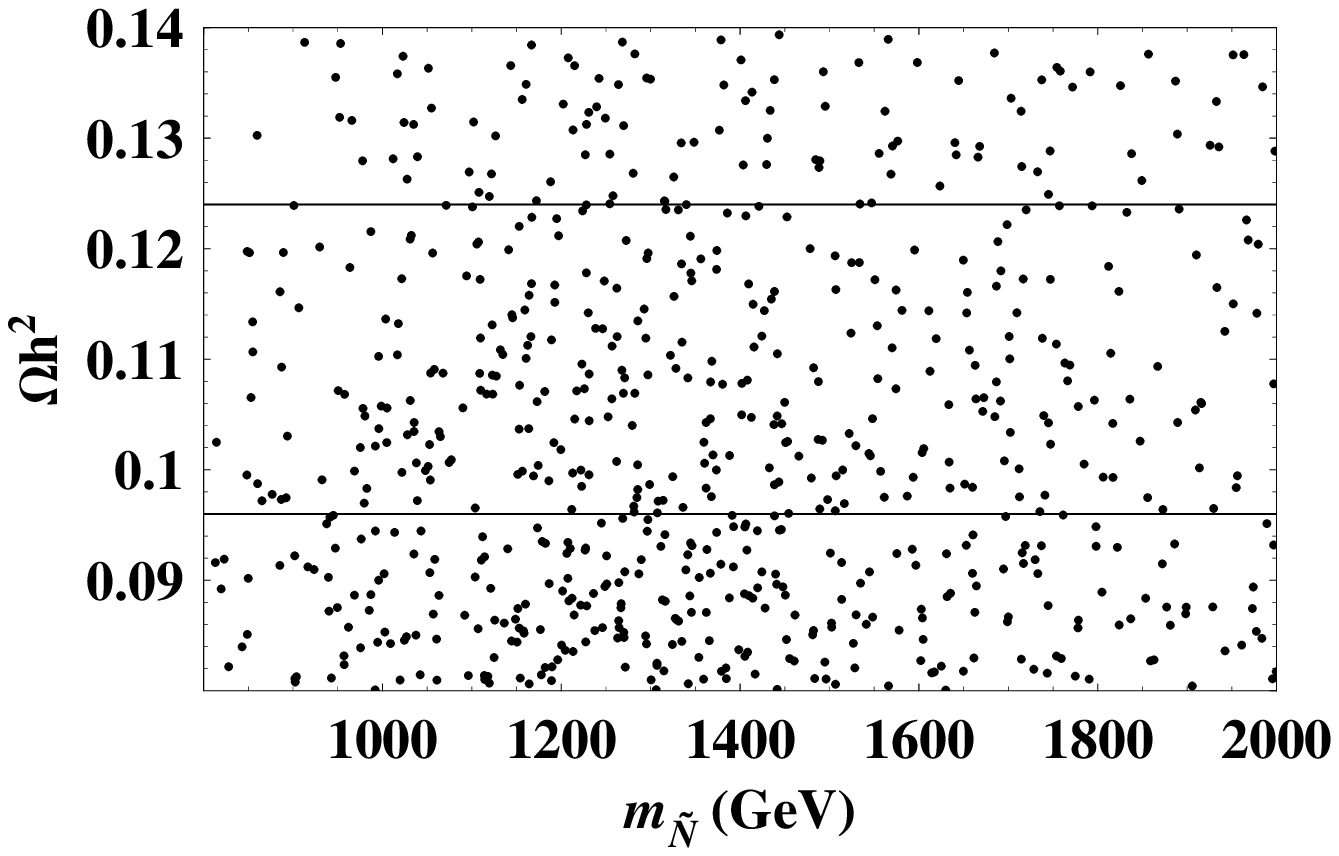}
\end{center}
\vskip -0.25in
\caption{We show the relic density as a function of the sneutrino mass for ${\tilde N} {\tilde N} \rightarrow NN$ (upper panel) and ${\tilde N}^* {\tilde N} \rightarrow \phi \phi$ annihilation channels dominated regions. The model points generated by varying the parameters mentioned in the text. The annihilation cross section is $\propto m^2_{Z'}$ in the latter case, which results in larger sneutrino masses. \label{ohsqvsm}}
\label{model}
\end{figure}

The sneutrino annihilation is dominated by two $S$-wave processes: ${\tilde N} {\tilde N} \rightarrow N N$ (${\tilde N}^* {\tilde N}^* \rightarrow N^* N^*$) and ${\tilde N} {\tilde N}^* \rightarrow \phi \phi$, where $\phi$ is the lightest Higgs in the $B-L$ sector. The neutrino final state is the main annihilation mode in most of the parameter space. In this case the relic density depends mostly on the mixing parameter for the new Higgses ($\mu'H_1'H_2'$) and is essentially independent from the $Z^{\prime}$ mass.
As a result, one can obtain the correct relic density for sneutrino mass as low as 60 GeV. We show the relic density in this model in Fig.~\ref{ohsqvsm} for $Z'$ mass of $1-2$~TeV. The $\phi$ final state is the main annihilation mode in regions of parameter space where $m_\phi \ll m_{\tilde N}$. The annihilation cross section in this case is $\propto m^2_{Z'}$~\cite{ADRS2}, which implies heavier sneutrinos for larger values of the $Z'$ mass. If $m_\phi < 20$~GeV, sneutrino annihilation at late times receives sufficient Sommerfeld enhancement and can explain the PAMELA results~\footnote{This can also be resolved without Sommerfeld enhancement if one invokes a non-thermal scenario where the sneutrinos are created from the decay of heavy moduli or gravitinos~\cite{Dutta:2009uf}.}~\cite{ADRS2}. The SUGRA models discussed in the previous two sections require an astrophysical boost factor $10^3-10^4$ to explain the PAMELA data by dark matter annihilation.

\begin{figure}[ht]
\begin{center}
\includegraphics[width=.48\textwidth]{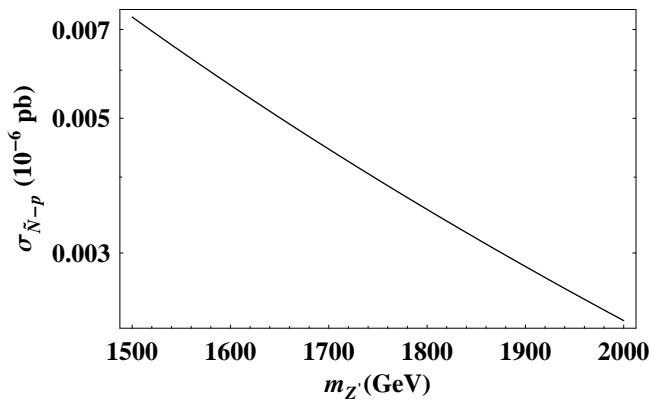}
\end{center}
\vskip -0.25in
\caption{We show the direct detection cross section in the $U(1)_{B-L}$ model as a function of $Z'$ mass.}
\label{directcdms}
\end{figure}

In the $B-L$ model the elastic scattering of the sneutrino off nucleons occurs via the $Z^{\prime}$ exchange in the $t$-channel. This leads to only a spin-independent contribution since the $B-L$ charges of the left and right quarks are the same. The sneutrino-proton cross section does not have errors due to the strangeness content of the proton, but the detection rate can still get uncertainties from the galactic velocity distribution.
The cross section for sneutrino-proton elastic scattering follows
\begin{equation} \label{sigma}
\sigma_{{\tilde N}-p} \propto \left({g_{B-L} Q_L \over {2 m_{Z^{\prime}}}}\right)^4 m^2_p ,
\end{equation}
where $g_{B-L}$ and $Q_L$ are the $U(1)_{B-L}$ gauge coupling and the normalized $B-L$ charge of leptons, respectively, and $m_p$ is the proton mass. We note that for $g_{B-L} \sim 0.4$, and with a normalization factor of $\sqrt{3/2}$, all of the gauge couplings unify at $M_{\rm GUT}$. The limits on the $Z^\prime$ mass from LEP and Tevatron are given by~\cite{tevatron,LEP},
\begin{equation} \label{Z'mass}
\frac{2 m_{Z^\prime}}{g_{B-L} Q_L} > 6 \; \mathrm{TeV} \;.
\end{equation}
This results in an upper limit on $\sigma_{{\tilde N}-p}$ of $\sim 7 \times 10^{-9} \;  \mathrm{pb}$. In Figure~\ref{directcdms}, we show the $Z^{\prime}$ mass as a function of the $\tilde N$-$p$ scattering cross section for this model. Note that this is independent from the sneutrino annihilation mode. If the CDMS finding holds in the future results then depending on the final value of the cross-section either this model will be found to be already ruled out by the LEP data or the new gauge boson $Z'$ will be found very soon at the LHC.

The whole plot in Fig.~\ref{directcdms} holds when we try to explain the PAMELA puzzle since $m_Z'\geq 1.5$ TeV allows us to satisfy the relic density using mostly ${\tilde N} {\tilde N}^* \rightarrow \phi \phi$ annihilation mode.
The sneutrino mass required to solve the puzzle is, however, $\geq 1$~TeV. At present, due to low statistics it is not possible to constrain the dark matter mass by the CDMS data, however in the future with more data we will be able to find out whether this model can explain the PAMELA results.

An interesting point is that in the $B-L$ model the number of muon events at IceCube from annihilation of sneutrinos captured in Sun only depends on  $m_{\tilde N}$ and $\sigma_{{\tilde N}-p}$. The reason being that the spin-dependent part of the cross section vanishes since the $B-L$ symmetry is vectorial. Hence, more data from the upcoming XENON 100 and CDMS experiments (with an LHC measurement of the dark matter mass), will pinpoint the expected muon events from the $B-L$ model at the IceCube. The largest number of muon events from Sun for the sneutrino mass in the $60-300$~GeV range is $58\, {\rm km}^{-2}{\rm yr}^{-1}$, which may be seen in 5 years of IceCube running~\cite{ABDR}~\footnote{
The muon event signal from annihilation in Earth
is too small to detect. However, with a modified velocity distribution, it is possible to raise the Earth event rate becomes $12\, {\rm km}^{-2}{\rm yr}^{-1}$ (the maximum Sun muon rate can become $78\, {\rm km}^{-2}{\rm yr}^{-1}$ in this case).}.

We note that in mSUGRA models the neutralino has a large capture rate only in the hyperbolic branch/focus point region due to a large Higgsino component that results in
a large spin-dependent scattering cross section via $Z$ exchange. The number of muon events from Sun is in the range $100-1000\, {\rm km}^{-2}{\rm yr}^{-1}$ for the neutralino mass in the $60-300$~GeV range. The nonuniversal Higgs masses models with smaller $\mu$ would show similar behavior. However, these models do not exhibit a direct relationship between the expected muon events at IceCube and the direct detection experiments.

\section{Conclusion}

In this paper we have considered the status of mSUGRA, nonuniversal Higgs masses and $U(1)_{B-L}$ models in light of the CDMS results. We summarize our findings:\\
\\
\noindent
{\bf (1)} A direct detection cross section of $10^{-9}-10^{-8}$~pb selects an interesting region of mSUGRA parameter space. For example, if we include $g_\mu-2$ data  in the analysis, then the $B_s\rightarrow\mu^- \mu^+$ decays might be observed in the upcoming results at the Fermilab for large values of $\tan\beta$. Also, the sparticle masses (especially in the stau coannihilation regions) and the gauginos and Higgsinos (in the hyperbolic branch/focus point regions)  are mostly within the reach of the LHC. For $\tan\beta=50$, the CDMS exclusion contour, we find, has almost become competitive with the $b\rightarrow s\gamma$ and $B_s\rightarrow\mu\mu$ bounds for the stau-neutralino coannihilation region and rules out most of the $g_\mu-2$ favored parameter space in the focus point region. For ${\rm tan}\beta=10$, however, the exclusion limit becomes competitive with the Higgs mass contour.  \\
\\
\noindent
{\bf (2)} The nonuniversal Higgs masses models allow a much larger region of $m_0$ and $m_{1/2}$ since the relic density can be satisfied by using a large Higgsino component or Higgs funnel. The larger direct detection cross section however selects smaller values of $\mu$, and $O(10^{-8}$~pb) fits well with intermediate values of $\mu$ around 500~GeV.
The larger Higgsino component in the hyperbolic branch/focus point of mSUGRA and the regions with smaller values of  $\mu$ in the nonuniversal Higgs model both give enhanced signal at the IceCube.  The CDMS exclusion line rules out large region of parameter space for smaller values of $\mu$.\\
\\
\noindent
{\bf (3)} The current $Z^{\prime}$ mass limits set an absolute upper bound $\sim 7 \times 10^{-9}$~pb on the cross section in the $B-L$ model. Therefore, if the CDMS results hold, a more precise determination of the cross section along with the new $Z^{\prime}$ searches at the LHC will decide the fate of this model.
Because of the absence of spin-dependent interactions, more data from the dark matter experiments on the scattering cross section and mass, and direct mass determination at the LHC,  will fix the predicted signal of the $B-L$ model at IceCube. That will also determine whether dark matter explanation of the PAMELA in the context of the $B-L$ model, that requires sneutrino masses larger than 800~GeV, is still viable.\\
\\
We note that any sensible mass determination is impossible from just two events ~\cite{talk}. Many more events are needed to determine the dark matter mass from the recoil spectrum due to the near degeneracy of the recoil spectrum for a wide range of masses (determination of mass from the recoil spectrum suffers also from the uncertainty of WIMP velocity
distribution and halo models) as discussed in Ref.~\cite{joel} and, therefore, there is no constraint on the dark matter mass at this point.

\begin{table}[tbp]
\center
\begin{tabular}{|c|c|c|c|c|}\hline
\multicolumn{2}{|c|}{Models}&$m_{\rm LSP}$(GeV)&$\sigma_{{\rm LSP}-p}$($\times
10^{-8}$pb)&Constraints\\\hline&$\tan\beta$&&&\\\cline{2-2}
&10&165&0.275&Higgs mass (114.4~GeV)\\
&10&143&0.45&Higgs mass (113.4~GeV)\\
mSUGRA (coannihilation region) &20&148&0.821&Higgs mass (114.4~GeV)\\
$A_0 = 0$ &20&126&1.6&Higgs mass (113.4~GeV)\\
&30&148&1.66&Higgs mass (114.4~GeV)\\
&30&124&4.5&Higgs mass (113.4~GeV) \& $b\rightarrow s\gamma$ \\
&40&148&3.43&$b\rightarrow s\gamma$\\
&50&181&3.00&$B_s\rightarrow\mu^-\mu^+$\\\hline
mSUGRA (low focus point region)& 25 & 66 & 6.5 &  \\
$A_0 = 0$&30&69&11 & (see caption) \\
&40&70.2& 23 &   \\
&50&70.2 & 114.2&  \\\hline
\multicolumn{2}{|c|}{B-L }&70&$0.7$& $Z^{\prime}$ mass limit \\\hline
\end{tabular}
\caption{Closest fits to $m_{\rm LSP}= 70$ GeV and $\sigma_{{\rm LSP}-p} = 3.8 \times 10^{-8}$ pb in the mSUGRA and $B-L$ models. The corresponding limiting constraint is given in each case. For the low focus point models, the constraints are the chargino mass bound, $b\rightarrow s\gamma$, and the Higgs mass bound (113.4~GeV). The situation for nonuniversal models is more relaxed than that in the mSUGRA case and is discussed in the text.
}
\label{data}
\end{table}

In table 1 we show closest fits for these models if the CDMS results $\sigma_{{\rm LSP}-p} \sim 3.8 \times 10^{-8}$~pb and $m_{\rm LSP} =70$ GeV (which is obtained using the maximum experimental sensitivity or the lowest point of the curve) are taken at face value.
We find that the stau-neutralino coannihilation region of the mSUGRA moodel cannot accommodate CDMS results. The problems are due to the Higgs mass constraint (for low ${\rm tan}\beta$) and
$b\rightarrow s\gamma$/$B_s\rightarrow\mu^-\mu^+$ constraints (for large ${\rm tan}\beta$). We show the results for Higgs masses of 113.4 GeV and 114.4 GeV, since the theoretical calculation can have $\sim 1$ GeV uncertainty. It is still possible to find a close fit in the focus point/hyperbolic branch region. For example, if we choose ${\rm tan}\beta=25$, $m_0=1150$ GeV and $m_{1/2}=200$ GeV. For this point the LSP is a mixture of Bino and Higgsino and the mass can be $m_{\rm LSP} \sim 70$ GeV. The $\sigma_{{\rm LSP}-p}$ is within a factor of 2 of the experimental value (the theoretical uncertainties due to nuclear form factors can be bigger than 2). Since $m_0$ is large, the constraints from $b\rightarrow s\gamma$ or $B_s\rightarrow\mu^-\mu^+$ do not apply. We find similar points for ${\rm tan}\beta=30,40,50$, but the cross section becomes larger. In the nonuniversal scalar mass models~\cite{non-U} it is easier to find a solution for these CDMS values, since the Higgs mass constraint can be lifted by choosing larger $A_0$ values and the $b\rightarrow s\gamma$ constraint can be relaxed by increasing the third generation sparticle masses. It is possible to have even smaller values of neutralino masses. For example, for ${\rm tan}\beta=10$, $m_{1/2}=200$ GeV, $m_0=500$ GeV, $A_0=1$ TeV, we can find $\sigma_{{\rm LSP}-p}\sim 4 \times 10^{-8}$ pb and $m_{\rm LSP} \sim 61$ GeV. For this point, the third generation universal sfermion mass is $m_0=830$~GeV to satisfy the $b\rightarrow s\gamma$ constraint.
Finally, the $B-L$ model (assuming that it does not explain the PAMELA excess) does not have any problem with 70~GeV mass for the dark matter, however the cross section, which is dictated by the $Z^{\prime}$ mass limit from LEP and Tevatron, is small by a factor of 6. This discrepancy could arise from the uncertainties in the detection rate due to galactic velocity distribution.

\section{Acknowledgements}

The work of B.D. is supported in part by the DOE grant DE-FG02-95ER40917. The work of Y.S. is supported in part by the DOE grant DE-FG02-04ER41308. We thank Rupak Mohapatra, Teruki Kamon and Joel Sander for useful discussions.


\end{document}